\documentclass[12pt,preprint]{aastex}
\begin{document}

\parindent=1.0cm

\title{HAFFNER 16: A YOUNG MOVING GROUP IN THE MAKING}

\author{T. J. Davidge}

\affil{Dominion Astrophysical Observatory,
\\National Research Council of Canada, 5071 West Saanich Road,
\\Victoria, BC Canada V9E 2E7}

\author{Eleazar R. Carrasco, Claudia Winge, Peter Pessev, Benoit Neichel, Fabrice Vidal}

\affil{Gemini Observatory/AURA, Southern Operations Center, Casilla 603, La Serena, Chile} 

\author{Francois Rigaut}

\affil{Australian National University, Research School of Astronomy and Astrophysics,
\\Mount Stromlo Observatory, Cotter Road,
\\Weston, ACT 2611, Australia}

\altaffiltext{1}{Based on observations obtained at the Gemini Observatory, which is 
operated by the Association of Universities for Research in Astronomy, Inc., under a 
cooperative agreement with the NSF on behalf of the Gemini partnership: the National 
Science Foundation (United States), the National Research Council (Canada), CONICYT 
(Chile), the Australian Research Council (Australia), Minist\'{e}rio da Ci\^{e}ncia, 
Tecnologia e Inova\c{c}\~{a}o (Brazil) and Ministerio de Ciencia, Tecnolog\'{i}a e 
Innovaci\'{o}n Productiva (Argentina).}

\begin{abstract}

	The photometric properties of main sequence (MS) and pre-main sequence 
(PMS) stars in the young cluster Haffner 16 are examined using images recorded with 
the Gemini South Adaptive Optics Imager (GSAOI) and corrected for atmospheric 
blurring by the Gemini Multi-Conjugate Adapative Optics System (GeMS). A rich 
population of PMS stars is identified, and comparisons with isochrones suggest 
an age $\lesssim 10$ Myr assuming a distance modulus of 13.5 (D = 5 kpc). 
This age is consistent with that estimated from the 
lower cut-off of the MS on the $K-$band luminosity function, and is $\sim 2$ Myr younger 
than the age found from bright MS stars at visible wavelengths. 
When compared with the solar neighborhood, Haffner 16 is roughly a factor 
of two deficient in objects with sub-solar masses. PMS objects in the cluster 
are also more uniformly distributed on the sky than bright MS stars. It is suggested 
that Haffner 16 is dynamically evolved, and that it is shedding 
protostars with sub-solar masses. Young low mass clusters 
like Haffner 16 are one possible source of PMS stars 
in the field. The cluster will probably evolve on time scales of $\sim 100 - 1000$ Myr
into a diffuse moving group with a mass function that is very different from that 
which prevailed early in its life.

\end{abstract}


\section{INTRODUCTION}

	The total mass of a star cluster plays a key role 
in defining the pace of its evolution. While cluster 
mass affects the time scale of dynamical evolution in a direct manner, it may also 
influence cluster evolution in more subtle ways. Feedback from 
massive stars may disrupt a cluster if substantial mass outflow is induced while 
the cluster is in a gas-rich, pre-virialized state (e.g. Smith et al. 2011). However, 
such feedback-driven mass loss may not occur early-on in low mass clusters as 
there is a low probability that massive hot stars will form in systems with stellar 
masses less than a few hundred times solar assuming a solar neighborhood mass function 
(MF). Clusters that lack stars with evolutionary timescales that are shorter than 
a few cluster crossing times may then not be subject to the catastrophic mass loss 
that is thought to spur `infant mortality' (Lada \& Lada 2003), although environmental 
mechanisms might still contribute to their early demise (e.g. Kruijssen et al. 2012). 

	Cluster mass may also influence stellar content. 
While the characteristic masses of stars are primarily 
defined by physical constants rather than environmental properties 
(Krumholz 2011), the global characteristics of the host cluster may still influence the 
observed MF. If a large-scale loss of gas does not occur early-on then a cluster 
may retain low mass stars that would otherwise be lost (Pelpussey \& Portegies Zwart 
2012) for a longer period of time. Radiation and winds from hot stars can also 
erode proto-stellar accretion disks, thereby choking growth and limiting the minimum 
mass of MS stars (e.g. De Marchi et al. 2011b). If 
very massive stars do not form then circumstellar accretion disks may survive for 
longer periods of time than in more massive clusters. The minimum 
stellar mass in low mass clusters may then differ from that in more massive systems.

	Low mass clusters are expected to form in large 
numbers, and so are significant contributors to the field 
star population. Studies of low mass clusters in the early stages of 
their evolution may then provide insights into the overall properties of field stars, 
such as the observed MF and the binary frequency (e.g. Marks \& Kroupa 2011).
Haffner 16 is an open cluster with a mass of a few hundred solar masses 
that is located at Galactic co-ordinates $\ell = {0\fdg5}$ and $b = {242\fdg1}$ 
(i.e. outside of the Solar circle). Vogt \& Moffat (1972) present a color-magnitude 
diagram (CMD) that shows bright main sequence (MS) stars indicative of a young age. 
McSwain \& Gies (2005) measure an age of log(t$_{yr}$) = 7.08 from multicolor CCD 
photometry, and assign a distance of 3.2 kpc, placing Haffner 16 in the Perseus Arm. They 
also find 19 B stars that may be cluster members, one of which is a possible Be star. 

	Narrow-band H$\alpha$ measurements discussed by 
McSwain \& Gies (2005) indicate that there are a number of 
emission line sources with a range of broad-band colors in 
and around Haffner 16. Given the young age of the cluster, this 
suggests that there may be a substantial population of 
actively accreting pre-main sequence (PMS) stars. In the present paper, broad and 
narrow-band near-infrared images recorded with the Gemini South Adaptive Optics Imager 
(GSAOI) and corrected for atmospheric blurring by the Gemini Multi-congugate 
Adaptive Optics (AO) System (GeMS) are used to examine the photometric 
properties and spatial distribution of stars and proto-stars in Haffner 16. The use 
of GeMS is of interest for such an investigation given the compact nature of Haffner 16, 
and the potential for crowding if low mass stars are present in moderately large numbers.

\section{OBSERVATIONS}

	The images were recorded with GSAOI$+$GEMS (McGregor et al. 
2004; Carrasco et al. 2012, Rigaut et al. 2012; Neichel et al. 2012) as part of 
program GS-2012B-SV-409. GSAOI is an imager that is designed to be 
used with GEMS. The GSAOI detector is a mosaic of 
four $2048 \times 2048$ Rockwell Hawaii-2RG arrays, deployed in a $2 \times 2$ 
format. A $85 \times 85$ arcsec$^2$ field is imaged with 0.02 arcsec per pixel 
sampling. There are $\sim 3$ arsec gaps between the arrays. 
Haffner 16 fits within a single GSAOI pointing.

	GeMS produces image profiles that are more spatially uniform than would be 
delivered by a classical single beacon AO system over a similarly-sized field (e.g. 
Davidge 2010). This is done using information provided by five 
laser guide stars (LGSs), three natural guide stars (NGSs), and 
an on-detector guider to monitor wavefront distortions. The three NGSs used for 
the Haffner 16 observations were selected based on their brightness and 
location. The NGSs more-or-less define the apexes of an equalatoral triangle-shaped 
asterism that spans most of the GSAOI science field. While the stability of the image 
profile and the level of correction delivered by GeMS depends 
on a number of factors, including guide star brightness, 
wind speed in the dominant turbulent layers etc, an asterism with this shape typically 
provides a reasonably stable point spread function (PSF) across most of the GSAOI field. 

	Images were recorded through $J$, $Ks$, and Br$\gamma$ filters \footnote[1]
{Various properties of these filters are described at 
http://www.gemini.edu/sciops/instruments/gsaoi/instrument-description/filters/}. 
Two different exposure times were used to broaden the magnitude range 
that was sampled. Three 5 sec exposures were recorded per filter to cover moderately 
bright cluster members, while $5 \times 30$ sec exposures in $J$ and $Ks$ and $5 
\times 60$ sec exposures in Br$\gamma$ were recorded to sample fainter 
stars. An on-sky dither pattern with a throw of a few arcsec along the east-west 
axis was employed to allow coverage in one of the gaps in the detector mosaic. 

	The initial processing of the data followed standard 
procedures for near-infrared images. Steps included 
dark subtraction, flat-fielding, and the subtraction of 
calibration frames that monitor thermal emission from warm objects along the optical 
path, such as dust on the GSAOI entrance window. The calibration images required for 
the last step were constructed by computing -- on a filter-by-filter basis -- 
the median signal at each pixel on the detector mosaic from all of 
the long exposure images recorded for the program. 
A mean sky level was subtracted from each image to correct for 
variations in sky brightness before computing the median. Since a number of clusters were 
observed for this program, and the data were recorded with a dither 
pattern, then the median signal at each pixel is more-or-less free of 
contributions from bright objects on the sky, and so monitors 
the thermal emission signatures that remain after the flat-field pattern is removed.

	The processed images in each filter were aligned to 
correct for the dither offsets. The final step before stacking the aligned images was to 
correct for distortions introduced by the system optics. These are most severe near 
the edge of the science field, and affect the data in two ways. First, the 
angular scale over which the distortion changes is a few tenths of an arcsec, 
and image combination is complicated if -- as with these observations -- dither offsets 
exceed these scales. Second, the distortions are chromatic, and so sources near 
the edge of the science field have different filter-to-filter locations if the 
individual images are aligned near the center of the science field. If left uncorrected, 
these distortions complicate the matching of photometric measurements recorded in 
different filters as well as the investigation of the spatial distribution of objects. 
The distortions were corrected in a differential way by mapping the 
$J$ and Br$\gamma$ images into the reference frame defined by the $Ks$ filter 
using the GEOMAP and GEOTRAN tasks in IRAF.

\section{PHOTOMETRIC MEASUREMENTS}

	Stellar brightnesses were measured with the PSF-fitting 
routine ALLSTAR (Stetson \& Harris 1988). The source catalogues, 
PSFs, and preliminary magnitudes that serve as input to ALLSTAR were obtained from 
tasks in the DAOPHOT (Stetson 1987) package. The PSF in each filter was constructed 
from 14 isolated, high S/N ratio stars located throughout the field. The photometric 
calibration was based on standard star observations that were made on the same 
night as the Haffner 16 observations, and the instrumental magnitudes in $Ks$ were 
transformed into $K$ magnitudes \footnote[2]{Throughout this paper $Ks$ is used to 
refer to the filter, while $K$ refers to the transformed magnitudes.}.

	The uniformity of the image profiles is investigated 
in Figure 1, which shows the FWHMs of the PSF stars; 
the final deep $Ks$ image is also shown. The FWHMs in the final 
Br$\gamma$ images are similar to those in $Ks$. 
Even though the data were recorded during relatively poor seeing conditions (85\%ile 
image quality; $\sim 0.8$ arcsec FWHM in $Ks$), the corrected images have 
FWHM $\leq 0.16$ arcsec in $Ks$, demonstrating that 
GeMS can deliver substantial improvements in image quality 
over the GSAOI field even during less-than-optimal conditions. The level of correction 
and spatial uniformity of the image profile degrades towards shorter 
wavelengths, as the size of the isoplanatic patch shrinks. 
This results in a larger mean FWHM and more variation in the FWHM in $J$ than 
in $Ks$. The uniformity of the PSF and the mean FWHM 
during typical seeing conditions will be much better than shown in Figure 1.

	The photometry was performed using spatially-variable PSFs to account for 
variations in FWHM. Because GeMS did a reasonable job of correcting the PSF 
across the GSAOI field, only a low-order fitting function was employed. 
A much more complicated PSF model would be required for 
AO images corrected with only a single beacon, even over angular scales that are only a 
fraction of the GSAOI field.

	Artificial star experiments were run to investigate sample completeness and 
uncertainties in the photometry. Artificial stars were generated using the 
empirical PSFs that were constructed from the final images, 
and were assigned colors that follow the main plume of points in the CMD. 
As with on-sky objects, artificial stars were only considered to be detected 
if they were recovered in both $J$ and $Ks$. The recovery 
statistics of artificial stars indicate that incompleteness 
becomes an issue when $K \gtrsim 18 - 18.5$. This matches the approximate magnitude 
in the CMD where the density of objects noticeably starts to drop towards fainter 
magnitudes (Section 4.1).

\section{CMDs AND LUMINOSITY FUNCTIONS}

\subsection{The $(K, J-K)$ CMD}

	The $(K, J-K)$ CMD of Haffner 16 is shown in the left hand panel of Figure 2. 
The green error bars indicate the $\pm 2\sigma$ dispersion in $J-K$ at 
$K = 17$ and 18 calculated from the artificial experiments, and it is 
encouraging that the predicted dispersion matches the observed 
width of the red plume centered near $J-K = 0.9$ at $K = 18$. 
The artificial star experiments indicate that the positional scatter 
between objects with $K \lesssim 17$ on the CMD is due to differences 
in their intrinsic properties, rather than uncertainties in the photometry. 

	There is the potential for substantial field star contamination 
given that Haffner 16 is at a low Galactic latitude. However, as the cluster just 
fits within the GSAOI science field, then the density of field 
stars can not be measured from these data alone. Insights into field star 
contamination were gleaned by examining the 2MASS $(K, J-K)$ CMD of an area 
near Haffner 16. While the 2MASS observations are substantially shallower 
than the GeMS images, the magnitude range sampled by 2MASS still overlaps with 
the region containing the brightest PMS objects in Haffner 16 (see below). 

	The brightnesses of objects in $J$ and $K$ image sections retrieved from 
the 2MASS archive were measured with ALLSTAR. The resulting $(K, J-K)$ CMD of 
objects that are immediately to the south of Haffner 16, with the photometric 
calibration set using zeropoint information in the 2MASS image headers, is shown 
in the left hand panel of Figure 3. The 2MASS field star CMD can be compared with the 
GeMS CMD of Haffner 16, which is shown in the right hand panel of the figure. 
A comparison of the two panels indicates that field stars are present 
in the color and magnitude ranges populated by objects in Haffner 16. However, 
the projected number density of field stars is much lower than that of cluster members. 
To demonstrate this, the number of field stars expected in various regions of the 
CMD in the GSAOI field are indicated in brackets in the right hand panel of the 
figure. The expected number of field stars along the cluster sequence is 
substantially lower than the number of objects in the GeMS CMD, indicating that the 
majority of objects in the Haffner 16 CMD that have $K < 15$ are not field stars. 

	The 2MASS data provide a useful check of the GeMS photometric calibration. The 
2MASS CMD of objects in the same field that is sampled with GeMS is shown in the right 
hand panel of Figure 3. There are substantial differences between the angular resolution 
and photometric depth of the 2MASS and GeMS datasets, and some of the stars in Haffner 
16 are almost certainly photometric variables. Still, the 
GeMS and 2MASS CMDs are very similar.

	The 2MASS images also can be used to gain insights into the total brightness 
and mass of Haffner 16. The distribution of stars in the 2MASS image indicates that 
Haffner 16 has a radius $\sim 1$ arcmin and a total brightness M$_K = -3.2$ if the 
distance modulus, $\mu_0$, is 12.50 and M$_K = -4.2$ if $\mu_0 = 13.50$. The Bressan et 
al. (2012) models predict that a $1 M_{\odot}$ SSP with an age near 10 Myr will have 
M$_K = 2.8$, and so the estimated total mass of Haffner 16 is $\sim 260$ M$_{\odot}$ if 
$\mu_0 = 12.50$, and 630 M$_{\odot}$ if $\mu_0 = 13.50$.

	The middle and right hand panels of Figure 2 show comparisons between 
the $(K, J-K)$ CMD of Haffner 16 and $Z = 0.016$ sequences 
from the Padova Isochrones (Bressan et al. 2012), which were downloaded from the Padova 
Observatory web site \footnote[3]{http://stev.oapd.inaf.it/cgi-bin/cmd}. 
These models include PMS evolution, making them useful 
for examining the stellar contents of young clusters like Haffner 16. 
The approximate magnitude of the main sequence cut-off (MSCO), which is the faint 
end of the MS defined by stars that are relaxing onto the MS, is marked for each model. 
We caution that there are numerous uncertainties in the input physics of the PMS phase 
of evolution (e.g. Seiss 2001). For example, the mechanics of accretion affects 
profoundly the luminosities of PMS models (e.g. Offner \& McKee 2011). 

	The distance and color excess measured by McSwain \& Gies (2005) have 
been used to place the isochrones in the middle panel of Figure 2.
That the isochrones in this panel agree with the near-vertical red plume of stars with 
$K > 16.5$ suggests that the reddening is reasonable. However, many of the sources in 
the upper half of the CMD, the vast majority of which are probably cluster 
members (e.g. the comparisons in Figure 3), fall blueward of the 
isochrones if $\mu_0 = 12.50$. The PMS stars have ages between 10 and 25 Myr with this 
distance modulus, and many probable PMS objects with $K$ between 15 and 15.5 fall 
blueward of the 25 Myr isochrone. 

	The poor agreement with the blue envelope on the CMD 
can be eased -- while still maintaining a good match with the lower portion of the CMD -- if a larger distance modulus is adopted. This is demonstrated in 
the right hand panel of Figure 2, where comparisons are made with isochrones assuming 
$\mu_0 = 13.5$. The group of points with $J-K \sim 0$ and $K$ between 15 
and 15.5 that fall blueward of the 25 Myr isochrone in the middle panel now 
fall near the 10 Myr isochrone, and thus have a predicted age that is in 
better agreement with that of other PMS stars in Haffner 16. 
A distance modulus of 13.5 is also favored given the 
presence of numerous red emission line sources (McSwain \& Gies 2005), which are 
probably PMS stars and would not be expected to show line emission if they were 
much older than $\sim 10$ Myr (e.g. de Marchi et al. 2011). Using the 10 Myr 
isochrone in the right hand panel of Figure 3 as a guide, the location of points 
on the CMD suggest that the MSCO occurs near $K \sim 15.4$.
The issue of distance aside, it is clear from Figure 2 that Haffner 16 
harbors a rich population of PMS stars, as expected given its young age.

\subsection{Br$\gamma$ emission}

	The comparisons with the isochrones suggest that the majority of 
bright objects in Haffner 16 are MS stars, and that PMS objects dominate 
at fainter magnitudes. If the PMS stars in Haffner 16 are still accreting gas, which 
might be expected if they have near-solar metallicities and ages $\leq 10$ Myr 
(e.g. De Marchi et al. 2011), then they may show neutral Hydrogen in emission. 
Not all PMS stars will show such emission, as the accretion rate 
varies with time (Haisch et al. 2001; Calvet et al. 2000; Fedele et al. 2010). The 
rate of accretion, and hence emission line strength, also depends on (1) metallicity, 
in the sense that at a given age the accretion rate increases as metallicity is lowered 
(e.g. de Marchi et al. 2011), as well as (2) age, with the rate of accretion 
among Galactic PMS objects dropping by 1.6 dex per decade in age 
(e.g. Figure 9 of de Marchi et al.  2011). 

	Br$\gamma$ falls within the wavelength range covered by GeMS, 
and Br$\gamma$ emission has been detected from protostars (e.g. 
Connelley \& Green 2009; Beck et al. 2010; Doneshaw \& Brittain 2011). 
Because it involves a higher excitation transition, Br$\gamma$ will 
not be as strong as H$\alpha$, and previous studies have measured equivalent widths for 
Br$\gamma \leq 10$ \AA. The effective width of the Br$\gamma$ filter is 320\AA, and 
so the detection of emission with an equivalent width of a few \AA\ requires 
photometry with a reliability of a few hundredths of a magnitude. 

	The $(K, Br\gamma-K)$ diagram of Haffner 16 is shown in the upper panel of 
Figure 4. The GSAOI $Ks$ and Br$\gamma$ filters have similar mean throughputs, 
and so the Br$\gamma$ measurements were calibrated 
by scaling the $K-$band zeropoint according to the effective wavelength ranges of 
the two filters so that m$_{Br\gamma} \approx$ m$_{Ks}$. 
There is considerable scatter in the $Br\gamma - K$ 
values; still, between $K = 13$ and 17 there is a tendency for $Br\gamma-K$ to 
decrease towards fainter $K$. Near $K = 17.25$ the $Br\gamma-K$ distribution 
changes, as objects with $Br\gamma-K \gtrsim 0.1$ appear in large numbers. 
A Kolmogorov-Smirnov test indicates that the 
$Br\gamma-K$ distributions in the intervals $K = 14.75$ to 17.25 and 
$K= 17.25$ to 18.25 differ at more than the 99\% confidence level.

	The $1\sigma$ dispersion in $Br\gamma-K$ at $K = 17$ calculated from the 
artificial star experiments is $\pm 0.04$ magnitudes, and this matches 
the width of the $Br\gamma-K$ distribution at various $K$ magnitudes 
in the top panel of Figure 4. While photometric
uncertainties of this size make it difficult to detect individual 
sources of Br$\gamma$ emission with equivalent widths of a few \AA, random 
uncertainties can be suppressed by examining mean Br$\gamma - K$ values. 
The mean values of Br$\gamma-K$ in $\pm 0.5$ magnitude intervals in $K$ are 
shown in the lower panel of Figure 4. An iterative $\pm 2.5\sigma$ rejection filter 
was applied to suppress outliers, and the errorbars show the formal $\pm 1\sigma$ 
uncertainties in the mean. $<Br\gamma-K>$ is not constant over the 
range of $K$ magnitudes investigated here, with $<Br\gamma-K>$ at $K = 16.5$ differing 
from the means at $K=15.5$ and 17.5 at almost the $3\sigma$ level. 
To further investigate the behaviour of Br$\gamma-K$ with magnitude, a 
Kolmogorov-Smirnov test was used to compare the Br$\gamma - K$ distributions of 
sources with $K$ between 13 and 15 with those of sources having $K$ between 15 and 17.
The two Br$\gamma-K$ distributions differ at roughly the 
99\% confidence level, confirming that $<Br\gamma-K>$ changes with magnitude. 

	The magnitude range in a system with an age of 10 Myr that the Padova 
isochrones predict will contain PMS sources is shown in the lower panel of 
Figure 4 for $\mu_0 = 12.5$ and 13.5. $<Br\gamma-K>$ is lowest in these 
magnitude ranges. While this is consistent with a large population of actively accreting 
PMS stars being present, we caution that the drop in $<Br\gamma - K>$ is not 
conclusive proof of line emission -- the lower values of $<Br\gamma-K>$ near $K = 15.5$ 
and 16.5 could also reflect a change in $Br\gamma$ {\it absorption} strength, as 
expected if MS stars with cooler temperatures grow in number towards fainter $K$ 
magnitudes. If the low $<Br\gamma-K>$ values near $K = 16$ and 17 are due to PMS 
objects then a spectroscopic survey of this cluster should reveal a large number of 
Br$\gamma$ emission sources, and this could be checked by obtaining $K-$band 
spectra using a multi-object spectrograph such as Flamingos 2.

	There is a prominent population of objects with 
$K \geq 17$ that have comparatively high $Br\gamma-K$ values. 
We suspect that these are not cluster members, and so are not part of the PMS 
population and do not mark a change in cluster content that 
might be associated with -- say -- the MSCO. Rather, these are probably field 
stars. The comparatively large $Br\gamma-K$ values of these objects suggest that they 
might have relatively deep Br$\gamma$ absorption, such as occurs in MS stars 
with mid-A spectral-type. MS stars with this spectral type and apparent 
magnitude would be located at a distance of 8 kpc, placing them in the outer disc of the 
Galaxy. In fact, there are objects with $J-K$ between 0.3 and 0.7 and $K$ between 17 and 
18 in the $(K, J-K)$ CMD that form what appears to be a reddened A star sequence. 
However, the $Br\gamma-K$ colors of this particular set of objects 
tend to be {\it smaller} than those of objects with redder 
$J-K$ colors. Alternatively, the objects with large $Br\gamma-K$ colors 
and $K \geq 17$ might be very cool intrinsically faint foreground dwarf stars that have 
deep molecular absorption features in the wavelength range covered by the Br$\gamma$ 
filter (e.g. Geballe et al. 2002).

\subsection{The Cluster Luminosity Function}

	The luminosity function (LF) of a cluster contains encoded information about 
its star-forming history (SFH) and the MF of its component stars. 
The ages of young clusters can be estimated from a change in source 
counts in the LF that occurs near the MSCO (e.g. Cignoni et al. 2010). 
The amplitude of this feature depends on a number of factors, including age, age 
dispersion, and the binning used to construct the LF. As demonstrated below, the Padova 
isochrones predict that in a simple stellar system with an age $\leq 10$ Myr this 
feature will appear as a discontinuity with an amplitude $\geq 0.2$ dex in a $K$ LF 
with 0.5 magnitude binning. Ages estimated from this feature are of interest since 
they do not rely on stars near the main sequence turn-off, the numbers of which 
in low mass clusters may be prone to stochastic effects. 

	The $K$ LF of Haffner 16 is shown in Figure 5. 
The LF has been corrected for field star contamination using 
star counts from the Robin et al. (2003) model Galaxy, which were obtained through 
the web interface \footnote[4]{http://model.obs-besancon.fr/}. A uniform absorption 
distribution of 0.15 mag/kpc was assumed, and experiments indicate that the model 
counts do not vary substantially if the amount of extinction is doubled. The full 
range of stellar types and ages allowed by the model were included. 
The star count models agree to within a few percent 
with number counts between $K = 13.5$ and 15.0 
obtained from 2MASS observations that sample the field near Haffner 16. 

	The LF tends to rise towards decreasing magnitudes, 
although there is a prominent notch near $K \sim 16$. 
Model LFs constructed from the Bressan et al. (2012) isochrones are shown in Figure 5, 
and these provide guidance for interpreting the observations. The models assume 
a Chabrier (2001) lognormal IMF and Z=0.016, and sequences with ages 6.5 and 10 Myr 
are shown. The models have been normalized to match the Haffner 16 number counts 
between $K = 16.25$ and 18.75, and results are shown for $\mu_0 = 12.5$ and 13.5.

	Given the magnitude interval used for normalization, both models 
more-or-less fit the faint end of the LF. However, there are subtle differences 
within the magnitude interval used for normalization. With 
$\mu_0 = 12.5$ the models predict a trend that is systematically shallower than 
that observed, while for $\mu_0 = 13.5$ the models are systematically steeper.

	Both the 6.5 and 10 Myr models tend to fall below the observations with 
$K \leq 15.5$ if $\mu_0 = 12.5$, and better agreement would not be 
achieved by considering models with older ages. Neither 
the 6.5 Myr nor the 10 Myr model predicts a drop in number counts near $K = 16$ 
if $\mu_0 = 12.5$. The 10 Myr model could be made to better match the 
number counts with $K \leq 15.5$ if a 0.35 dex shift upwards were applied, 
although this would then produce a discrepancy between the models and observations 
when $K \geq 16.5$, in the sense that Haffner 16 would be deficient in lower mass 
PMS objects when compared with model predictions.

	There is slightly better agreement with the observations 
if $\mu_0 = 13.5$. In this case the 10 Myr model matches the location -- if not the 
amplitude -- of the drop in number counts near $K = 16$. 
However, as with $\mu_0 = 12.5$, there is a discrepancy with the 
number counts at the bright end, in the sense that the agreement 
with the observed LF in the interval $K \leq 15.5$ could be improved by  
moving the models up by $\sim 0.25$ dex. There would then be a 
discrepancy between the models and observations at the faint end, in the 
sense that Haffner 16 would be deficient in low mass PMS objects when compared 
with model predictions.

	The ratio of the numbers of stars with $K \leq 15.5$ and $K \geq 16.5$ 
is smaller than predicted by the model LFs for both distance moduli.
This is not due to incompleteness, which only becomes significant for $K \geq 18.5$. 
Rather, to the extent that the models faithfully reproduce PMS evolution, then 
the comparisons in Figure 5 suggest that the MF in Haffner 16 is shallower than the 
Chabrier (2001) solar neighborhood relation. Later in the paper we will argue that 
this is likely a result of dynamical evolution.

\section{THE SPATIAL DISTRIBUTION OF MS AND PMS OBJECTS}

	The spatial distribution of objects forms part of a cluster's fossil record. 
The angular distributions of two groups of objects in Haffner 16 are investigated 
in this section: (1) bright main sequence (BMS) stars, which are defined to have $K$ 
between 13 and 15, and (2) PMS sources, which have $K$ between 15 and 17. 
For the purposes of defining these samples it has been assumed that $K = 15$ is the 
boundary between MS and PMS objects, based on the peak magnitude of the 10 Myr 
PMS sequence if $\mu_0 = 13.5$. The results presented below would not change markedly
if the BMS/PMS dividing point was set at -- say -- $K = 15.5$. 

	The on-sky locations of the sources in the BMS and PMS samples are compared in 
Figure 6. There are differences in the distributions of the BMS and PMS 
samples. While the PMS sources are more-or-less uniformly distributed, BMS objects tend 
to avoid the peripheral areas of the GSAOI field, suggesting that they are more centrally 
concentrated. 

	The two-point correlation function (TPCF) is one means of examining 
the clustering properties of sources. The TPCF is generated by computing the separations 
between all possible pairings in a sample, and then dividing the resulting 
separation function (SF) by that of a set of randomly distributed 
sources located in the same detector geometry (e.g. including gaps between arrays) as 
the source data. The TPCFs of the BMS and PMS samples, normalized using the SF of 
$10^4$ randomly distributed sources and scaled according to the number of 
pairings in the real and artificial datasets, are compared in the top panel of Figure 7.

	There is considerable bin-to-bin chatter in the TPCFs due to the small numbers 
of objects in the BMS and PMS samples. Still, the TPCFs of the BMS and PMS samples 
differ in a systematic way, in the sense that the TPCF of PMS objects is 
flatter than the TPCF of BMS objects. The PMS TPCF falls above 
that of BMS stars at separations $> 65$ arcsec, and below that 
of BMS stars for separations $< 30$ arcsec. There is also a modest peak in the TPCF 
of BMS objects in the bin that samples the smallest 
separations. This bin samples the angular scale where binaries, 
which have separations $\leq 0.04$ parsec (Larson 1995), might be expected 
at the distance of Haffner 16. However, results presented later in 
this section suggest that this feature is not statistically significant.

	The BMS and PMS samples have different levels of field star contamination. The 
Robin et al. (2003) model Galaxy predicts $\sim 7$ field stars in the BMS sample (out of 
27 objects total), and 29 in the PMS sample (out of 58 objects total). 
The field star fraction in the PMS sample is thus $\sim 50\%$. To assess biases in 
the TPCF that might arise from field stars, simulations were run in which 
a population of objects with locations drawn from a uniform on-sky field star 
distribution were added to the BMS sample. The number of objects added to the 
BMS sample was selected so that the fraction of field stars was the same as in the PMS 
sample. The SF and TPCF of this BMS$+$Field star sample was then computed. 
This was repeated using different samples of randomly distributed stars, and 
the mean TPCF of the entire suite of realizations is compared with the PMS TPCF in 
the bottom panel of Figure 7. The error bars show the $1\sigma$ standard deviations 
about the mean signal at each separation.

	The addition of objects with a uniform distribution on the sky introduces 
systematic trends in the BMS TPCF at separations $> 65$ arcsec, 
elevating the TPCF at these separations. Even so, 
the agreement with the signal in the PMS sample at large separations is poor. 
In addition, the TPCF of the mean BMS$+$Field sample consistently falls above that of 
the PMS TPCF at separations 5 -- 20 arcsec, indicating that the PMS objects 
show a lower degree of clustering on these angular scales than the BMS stars. We 
conclude that the objects in the PMS sample are 
more uniformly distributed than the BMS stars.
Finally, there is considerable scatter in the BMS$+$Field star TPCF 
at the smallest separations, with no peak in the mean TPCF of BMS$+$Field objects 
due to close binaries. This indicates that the peak in the PMS sample in the 
smallest separation bin noted earlier is not a statistically robust feature.

\section{SUMMARY \& DISCUSSION}

	Near-infrared images obtained with GSAOI and GeMS on Gemini South have been 
used to investigate the stellar content of the young open cluster Haffner 16. Two 
distance moduli are considered: $\mu_0 = 12.50$ and $\mu_0 = 13.50$. The former is that 
measured by McSwain \& Gies (2005) while the latter is found to produce better 
agreement with the $(K, J-K)$ CMD at the bright end and better internal consistency 
among the ages of PMS stars, which are present in large numbers in Haffner 16. 

	Isochrones predict that the PMS stars in Haffner 16 have ages $\lesssim 10$ Myr 
if $\mu_0 = 13.5$, and so some of these may show Hydrogen line emission. In fact, the 
$y-H\alpha$ colors of sources with $b-y > 0.1$ in Figure 12 of McSwain \& Gies (2005) 
indicate that a rich population of H$\alpha$ emission sources are present in and around 
the cluster. McSwain \& Gies (2005) argue that at least some of the emission line objects 
with blue colors may be Be stars, and we suspect that a substantial fraction 
of the faint, red H$\alpha$ sources in their data are PMS objects. The 
$<Br\gamma - K>$ measurements discussed in Section 4.2 are also consistent 
with a significant fraction of the intermediate brightness PMS candidates 
in Haffner 16 having Br$\gamma$ in emission. Accretion activity might be 
expected to be long-lived in low mass clusters that did not form massive hot stars, 
as the disruption of accretion disks by feedback will be delayed. Velocity measurements 
would allow the membership status of the emission line objects to be assessed. 

	Comparisons with solar metallicity models suggest an 
age of $\sim 10$ Myr from the MSCO if $\mu_0 = 13.50$, 
and this is only a few Myr younger than the age estimated by 
McSwain \& Gies (2005) from bright MS stars (albeit assuming a 
different distance modulus). A difference in ages measured from bright MS stars and 
the MSCO might be expected, especially for a low mass cluster like Haffner 16. 
Ages that are based on the properties of the brightest cluster stars are 
subject to stochastic effects, in the sense that the objects with the shortest lifetimes, 
such as very massive MS stars or stars in advanced stages of evolution, may 
be missing. The absence of such stars, whether due to stochastic effects or 
limitations imposed by the masses of star-forming clumps (e.g. Oey 2011), will skew ages 
estimated from the location of the MS turn-off to larger values. 
An age dispersion of a few Myr could also contribute to differences between 
age measurements made from MS and PMS stars in very young clusters.
Star formation in at least some clusters has been found 
to occur over a 2 - 3 Myr time period (e.g. Hosokawa et al. 2011; Delgado et al. 2011; 
Bik et al. 2012, and references therein), while in giant star-forming complexes 
(e.g. De Marchi et al. 2011a) in-falling gas may fuel star-formation for 
even more extended periods of time (e.g. Garcia-Benito et al. 2011). 

	The data discussed here provide insights into the dynamical state of Haffner 16, 
and there are indications that it is dynamically evolved. To the extent that the 
Padova models faithfully track PMS evolution, the comparisons in Section 4 
indicate that PMS stars fainter than $K = 16.5$ occur in numbers that are lower than 
expected if stars in the cluster followed a Chabrier (2001) lognormal MF; Haffner 16 
is deficient by roughly a factor of two in sub-solar mass PMS objects when compared with 
the solar neighborhood, and this holds for both distance moduli considered here. 
Also, evidence is presented in Section 5 that intermediate brightness PMS 
objects -- which have sub-solar masses in Haffner 16 -- are less clustered than 
the more massive MS stars. A trend of higher separations 
among PMS stars has been seen in other young clusters 
(e.g. Delgado et al. 2011), and is a signature of mass segregation.

	The dynamical state of Haffner 16 can be assessed by comparing its 
age and dynamical timescale. The dynamical timescale is $\sim 1.1 - 1.7$ Myr using the 
radius and mass estimates computed from the 2MASS data in Section 4; thus, the 
cluster has a dynamical age of 6 - 9 crossing times. We have detected $\sim 150$ 
objects in Haffner 16, and an extrapolation of the LF 
to $K = 19 - 20$, which is the expected brightness of a 0.10 M$_{\odot}$ proto-star 
with an age 10 Myr based on the Bressan et al. (2012) models, suggests that $\sim 100$ 
stars are below the faint limit if $\mu_0 = 12.50$, while 250 objects are missed if 
$\mu_0 = 13.5$. The ratio of relaxation to core-crossing time scales is then (Equation 
4--9 of Binney \& Tremaine 1987) $\sim 6\ (\mu_0 = 12.5) - 8\ (\mu_0 = 13.5)$. 
The time scale for mass segregation is shorter than that for system relaxation, 
and so the detection of mass segregation signatures in 
Haffner 16 is then not surprising. A culling of low mass 
objects may also occur during early epochs if there is large scale gas removal 
(Pelupessy \& Portegies Zwart 2012).

	If it is assumed that the IMF of the Galactic thin 
disk is universal, then Haffner 16 has shed a significant fraction of its PMS stars. 
A diffuse collection of PMS stars might then be expected outside of 
the area probed with GeMS. Compact low mass clusters like Haffner 16 that have 
relatively short dynamical timescales may be a source of isolated groupings 
of PMS stars like those seen in the Galactic disk (e.g. Webb et al. 1999).
Haffner 16 will almost certainly become progressively more diffuse 
on a timescale of $\sim 100$ Myr, eventually resembling a loosely concentrated 
moving group populated by a remnant ensemble of intermediate age 
stars with a MF that is very different from that with which it was born. 

\acknowledgements{It is a pleasure to thank the anonymous referee for providing a 
timely and comprehensive report.}

\clearpage

\begin{figure}
\figurenum{1}
\epsscale{1.00}
\plotone{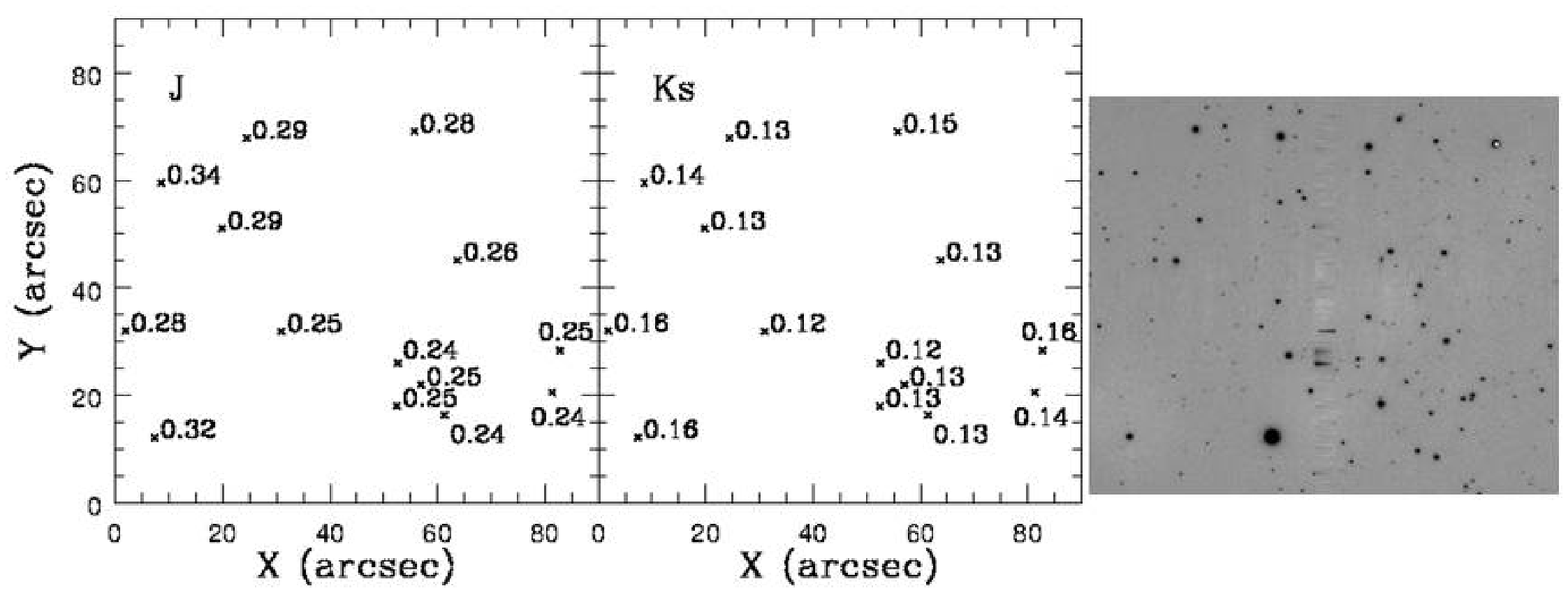}
\caption{Full-width at half maximum (FWHM) measurements of the 
PSF stars. The FWHMs are in arcsec, while $X$ 
and $Y$ are angular offsets in the GSAOI field measured from one 
corner of the mosaic. The final deep $Ks$ image is shown in the 
right hand panel. The mean FWHM in $Ks$ is significantly 
smaller, and has much better spatial uniformity than in $J$.}
\end{figure}

\clearpage

\begin{figure}
\figurenum{2}
\epsscale{0.75}
\plotone{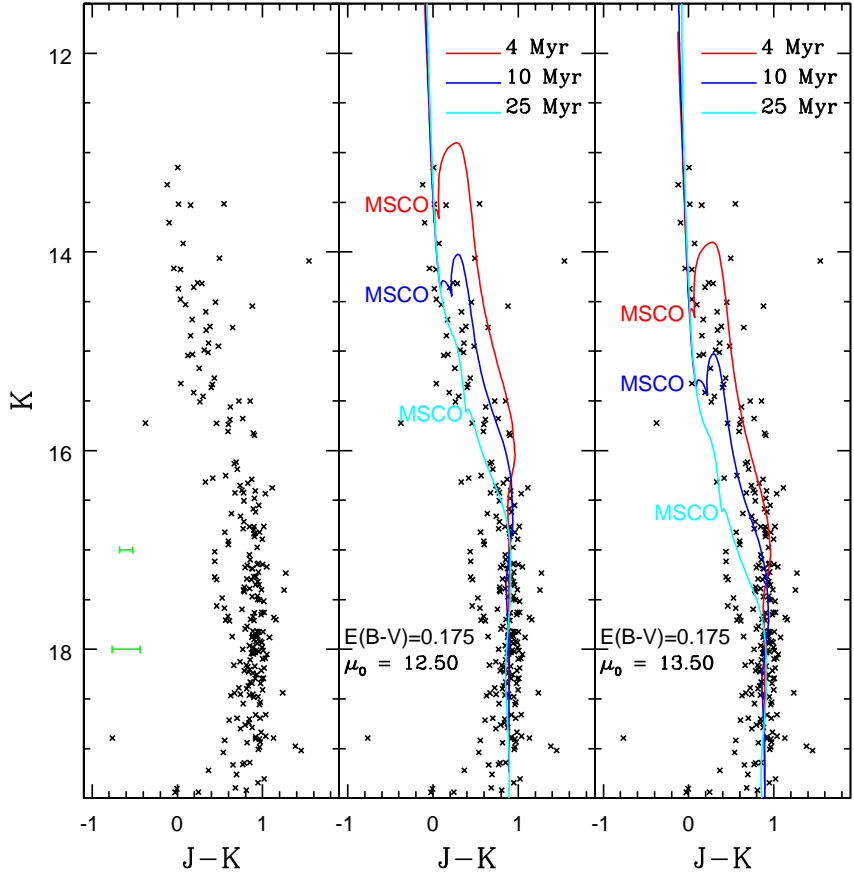}
\caption{The $(K, J-K)$ CMD of Haffner 16 is shown in the left hand panel, where 
the green error bars indicate the $\pm 2\sigma$ dispersion in $J-K$ measured from 
artificial star experiments. The dispersion in $J-K$ at $K = 18$ calculated from the 
artificial star experiments matches the width of the red plume near $J-K = 0.9$. 
Comparisons with $Z = 0.016$ sequences from the Padova Isochrones (Bressan et 
al. 2012) are made in the middle and right hand panels. The magnitude 
of the main sequence cut-off (MSCO), which is the faint limit of 
stars that have contracted onto the main sequence, is indicated for each isochrone. 
The models in the middle panel assume the reddening and distance modulus 
measured by McSwain \& Gies (2005). While a number of objects fall blueward 
of the isochrones when $K \leq 17$, there is reasonable agreement with the vertical 
plume of objects with $J-K \sim 0.9$ near the faint end of the CMD, 
suggesting that the adopted reddening is reasonable. The result of adopting the same 
reddening but a larger distance modulus is examined in 
the right hand panel. A distance modulus of 13.5 produces better agreement with 
the blue envelope of the CMD, and the majority of PMS stars at intermediate 
brightnesses now fall between the 4 and 10 Myr isochrones. 
The distance modulus notwithstanding, it is evident that 
Haffner 16 contains a large population of PMS stars.}
\end{figure}

\clearpage

\begin{figure}
\figurenum{3}
\epsscale{0.75}
\plotone{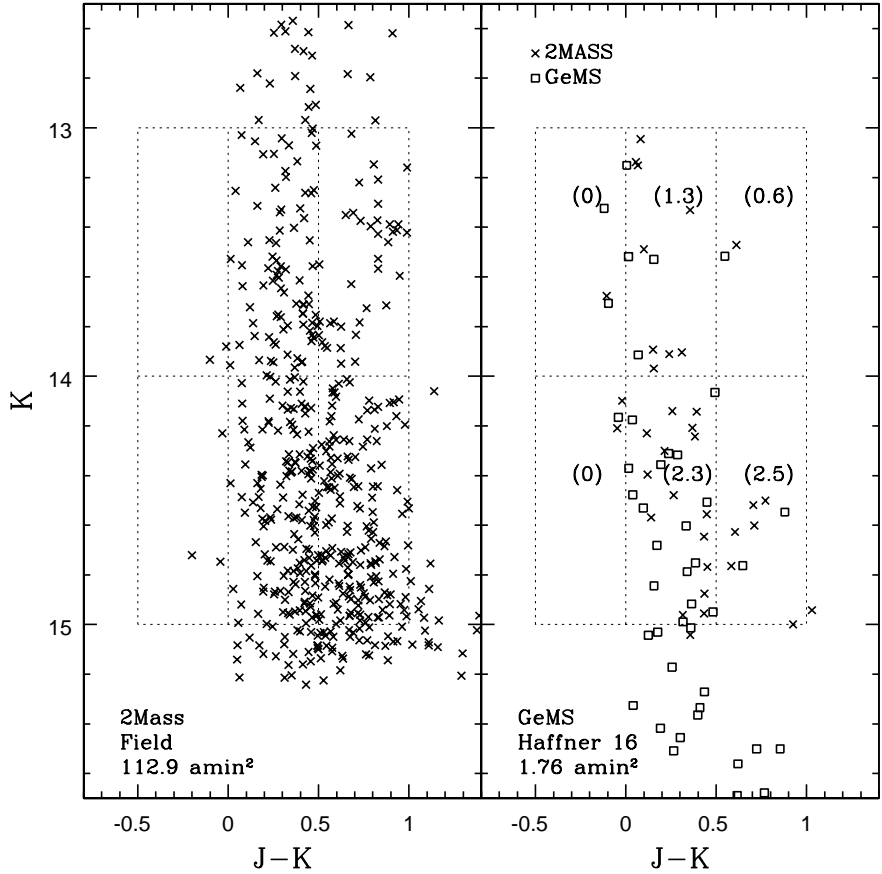}
\caption{The 2MASS $(K, J-K)$ CMD of a 112.9 arcmin$^2$ area 
that is immediately to the south of Haffner 16 is shown in 
the left hand panel, while the GeMS $(K, J-K)$ CMD of Haffner 16 is plotted in 
the right hand panel. The numbers of field stars expected in various areas of the 
GeMS CMD, based on counts in the 2MASS CMD in the left hand panel, are indicated in 
brackets. These numbers indicate that the vast majority of objects with $K < 
15$ in the GeMS CMD are not field stars. Also plotted in the right hand panel 
is the 2MASS CMD of Haffner 16, sampling the same area on the sky as GSAOI. 
The differences in angular resolution and photometric depth between the 2MASS and GeMS 
datasets notwithstanding, the two CMDs are very similar.}
\end{figure}

\clearpage

\begin{figure}
\figurenum{4}
\epsscale{0.75}
\plotone{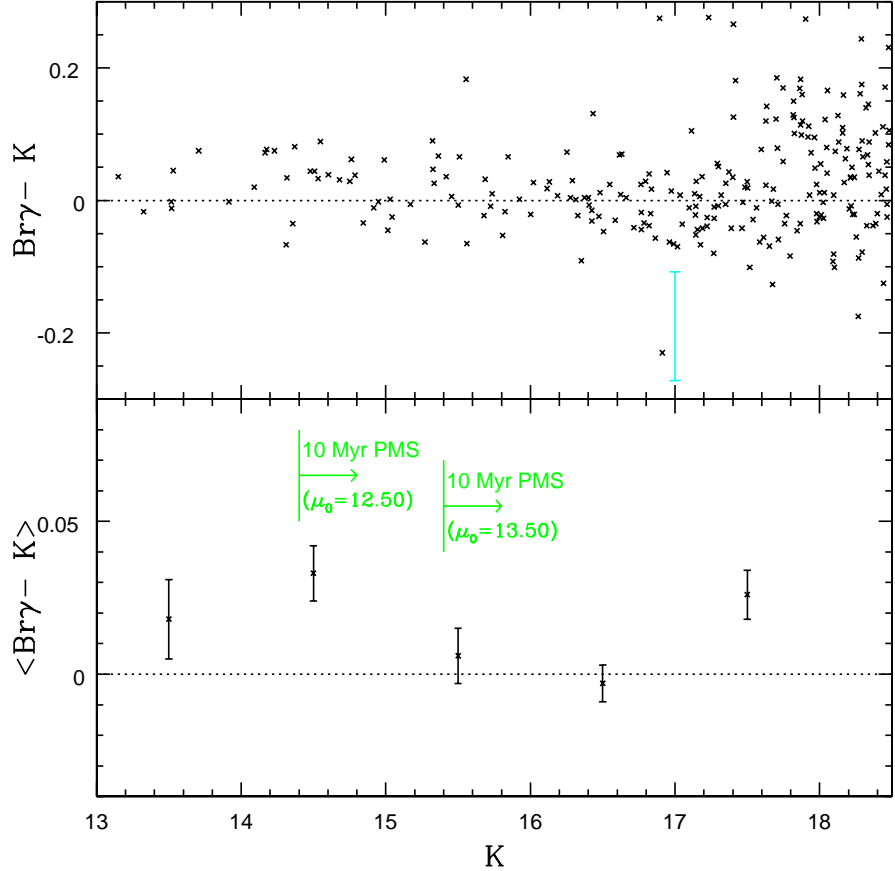}
\caption{The $Br\gamma-K$ colors of individual sources is shown in the top panel, while 
the mean $Br\gamma-K$ colors in $\pm 0.5$ magnitude $K$ intervals are shown in the 
bottom panel. Br$\gamma$ emission will cause objects to have lower Br$\gamma - K$ colors 
than non-emitting sources. The cyan errorbar in the top panel is the $\pm 2\sigma$ 
dispersion in $Br\gamma-K$ at $K = 17$ calculated from the artificial star experiments, 
while the errorbars in the lower panel are the $1\sigma$ uncertainties in 
$<Br\gamma-K>$. $Br\gamma-K = 0$ is indicated in each panel. The magnitude range where 
the Padova isochrones predict PMS stars should occur in a system with an age of 10 
Myr is indicated in the lower panel for $\mu_0 = 12.50$ and 13.50. $<Br\gamma-K>$ 
is not constant, but changes with $K$, with $<Br\gamma-K>$ being lowest 
in the magnitude range where PMS stars are expected. $<Br\gamma-K>$ increases 
when $K \geq 17$, and a comparison with the top panel indicates that this is due at least 
in part to the onset of a population of objects with $Br\gamma-K \gtrsim 0.1$.}
\end{figure}

\clearpage

\begin{figure}
\figurenum{5}
\epsscale{0.75}
\plotone{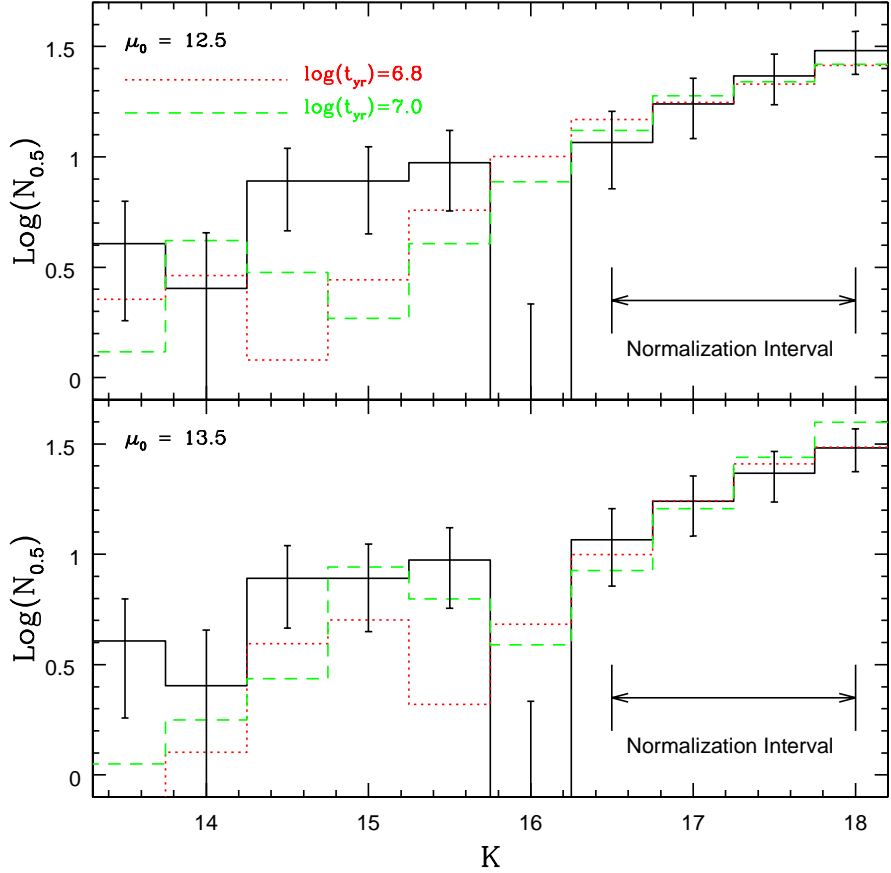}
\caption{The $K$ LF of Haffner 16. $N_{0.5}$ is 
the number of objects per 0.5 magnitude interval in $K$, corrected for field 
star contamination using results from the Robin et al. (2003) model Galaxy. The 
error bars are $1\sigma$ uncertainties based on counting statistics. Synthetic 
LFs constructed from Z=0.016 Bressan et al. (2012) models, with 
ages of log(t$_{yr}$) = 6.8 (dotted line) and 7.0 (dashed line), 
are also shown. The models assume a Chabrier (2001) lognormal 
IMF, and have been scaled to match the observations in the magnitude interval 
indicated. Comparisons are shown for two distance moduli. The drop in the 
number counts near $K = 16$ can be reproduced -- at least in location if not in 
amplitude -- by the 10 Myr models with $\mu_0 = 13.5$. The models tend to 
fall below the observations when $K \leq 15.5$ for both distance moduli. 
This suggests that the MF of Haffner 16 is shallower than 
predicted by the models, and it is argued that this is likely a consequence of 
dynamical evolution.}
\end{figure}

\clearpage

\begin{figure}
\figurenum{6}
\epsscale{0.75}
\plotone{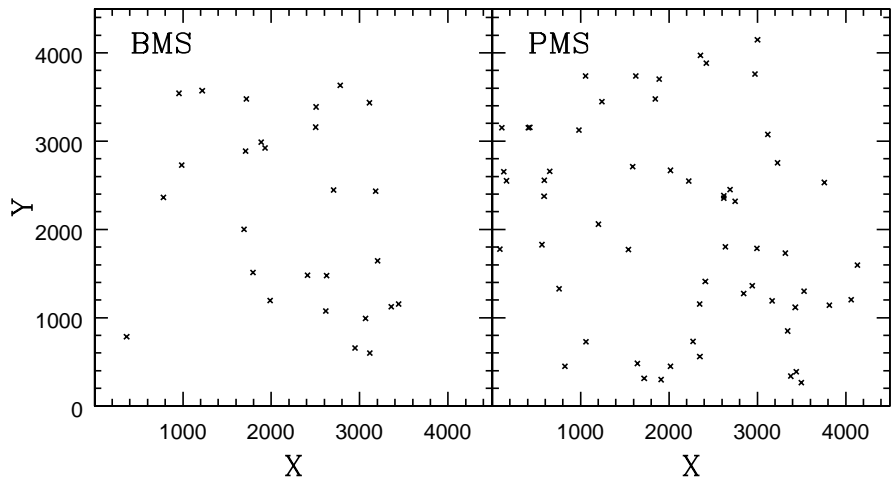}
\caption{The spatial distribution of objects in the BMS and PMS samples. $X$ and 
$Y$ are GSAOI pixel locations in the co-ordinate system defined by the $Ks$ observations 
(Section 2). Note that while the objects in the PMS sample are 
distributed more-or-less uniformly across the field, 
the BMS stars appear to be more centrally concentrated.}
\end{figure}

\clearpage

\begin{figure}
\figurenum{7}
\epsscale{0.75}
\plotone{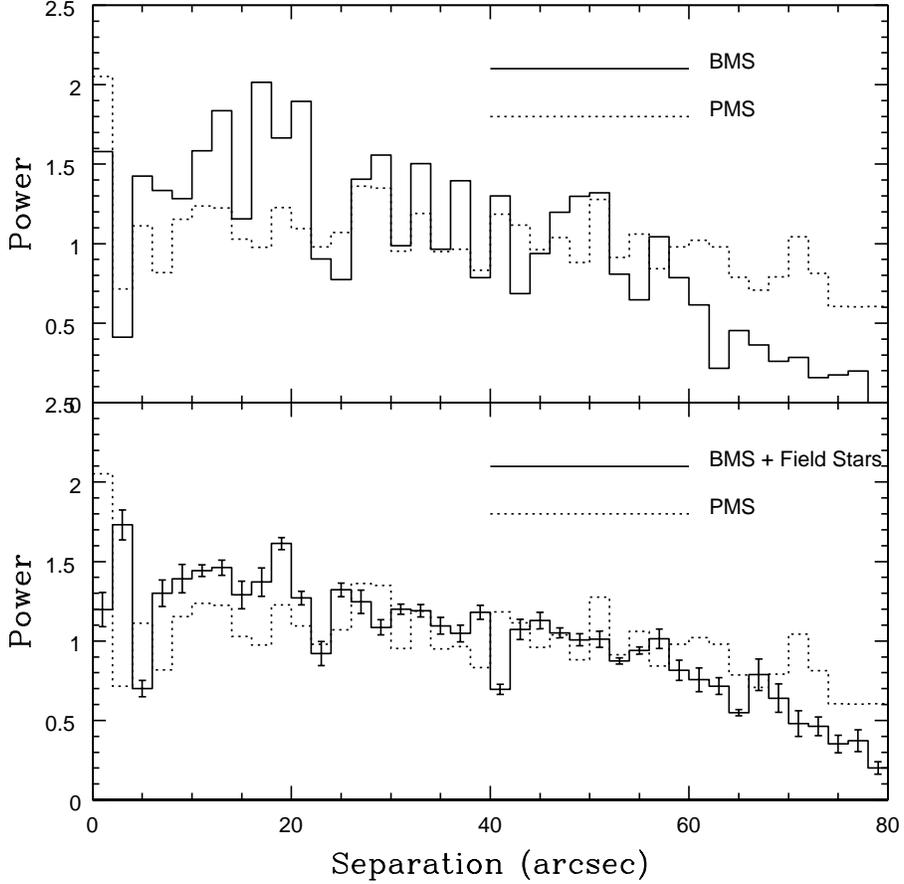}
\caption{Top panel: The two-point correlations functions (TPCFs) of the BMS and PMS 
samples. The separation functions (SFs) of both samples have been divided by the SF of 
$10^4$ randomly distributed sources, and then scaled according to the number of objects 
in the real and simulated samples. The TPCFs of the BMS and PMS sample differ, in that 
the PMS objects tend to have more power at separations $> 60$ arcsec, and less power 
at separations $< 20$ arcsec. Bottom panel: The solid line shows the average 
TPCF of a suite of simulations in which random objects, selected from 
a uniformly distributed population, were added to the 
BMS sample to simulate the 50\% field star contamination in 
the PMS sample. The error bars show the $1\sigma$ standard deviation about 
the mean TPCF at each separation. Also shown is the TPCF of the PMS sample from 
the top panel. Differences remain between the TPCFs of the BMS and PMS samples 
after compensating for differences in field star fraction, 
in the sense that the TPCF of the PMS sample is flatter than that 
of the BMS$+$Field star sample. These experiments confirm that PMS stars 
in Haffner 16 are more loosely clustered than BMS stars.}
\end{figure}

\end{document}